\documentclass[preprint,aps,prl,showpacs,twocolumn,mamsmath,10pt]{revtex4}
\usepackage{epsfig,epsf,graphics,psfrag}
\usepackage{times}
\usepackage{float}
\usepackage{color}
\usepackage{amsfonts,amssymb,stmaryrd,latexsym,amsmath}
\usepackage{hhline}

\newcommand{\be}{\begin{equation}}
\newcommand{\ee}{\end{equation}}
\newcommand{\ba}{\begin{eqnarray}}
\newcommand{\ea}{\end{eqnarray}}
\newcommand{\non}{\nonumber}

\begin{document}
\preprint{\small FZJ-IKP-TH-2009-22, HISKP-TH-09/23}
\title{Extraction of the light quark mass ratio from the
decays {\boldmath$\psi' \to J/\psi\pi^0(\eta)$}}
\author{Feng-Kun Guo$^1$
, Christoph~Hanhart$^{1,2}$
, and Ulf-G. Mei{\ss}ner$^{1,2,3}$
}

\affiliation{$\rm ^1$Institut f\"{u}r Kernphysik and J\"ulich Center
             for Hadron Physics, Forschungszentrum J\"{u}lich,
             D--52425 J\"{u}lich, Germany}%
\affiliation{$\rm ^2$Institute for Advanced Simulation,
             Forschungszentrum J\"{u}lich, D--52425 J\"{u}lich, Germany}
\affiliation{$\rm ^3$Helmholtz-Institut f\"ur Strahlen- und
             Kernphysik and Bethe Center for Theoretical Physics,\\ Universit\"at
             Bonn,  D--53115 Bonn, Germany}

\begin{abstract}
\noindent
 Light quark masses are important fundamental
 parameters of the Standard Model. The decays
 $\psi'\to J/\psi\pi^0(\eta)$ were widely used in determining
 the light quark mass ratio $m_u/m_d$. However, there is a large discrepancy
 between the resulting value of $m_u/m_d$ and the one determined from
 the light pseudoscalar meson masses. Using the technique
 of non-relativistic effective field theory, we show that
 intermediate charmed meson loops lead to a sizeable
 contribution to the decays and hence make the $\psi'\to J/\psi\pi^0(\eta)$
 decays not suitable for a precise extraction of the light quark mass ratio.
\end{abstract}
\pacs{14.65.Bt, 13.25.Gv, 12.39.Hg}

\maketitle

The decays of $\psi'$ into $J/\psi\pi^0$ and $J/\psi\eta$ were
suggested to be a reliable source for extracting the light quark
mass ratio  $m_u/m_d$~\cite{Ioffe:1979rv,Ioffe:1980mx} (for reviews,
see Refs.~\cite{Donoghue:1989sj,Meissner:1993ah,Leutwyler:1996eq}).
The decay $\psi'\to J/\psi\pi^0$ violates isospin symmetry. Both the
up-down quark mass difference and the electromagnetic (em)
interaction can contribute to isospin breaking. However, it was
shown that the em contribution to the decay $\psi'\to J/\psi\pi^0$
is much smaller than the effect of the quark mass
difference~\cite{Donoghue:1985vp,Maltman:1990mp}. Based on the QCD
multipole expansion and the axial anomaly, the relation between the
quark mass ratio
\be%
{1\over R}\equiv{m_d-m_u\over m_s-\hat{m}},\label{eq:R}
\ee%
where $\hat{m}=(m_d+m_u)/2$, and the ratio of the decay widths of
these two decays was worked out up to the next-to-leading order in the
chiral expansion~\cite{Donoghue:1992ac,Donoghue:1993ha}. At
leading order, the relation reads~\cite{Leutwyler:1996qg}
\be%
R_{\pi^0/\eta} \equiv \frac{{\cal B}(\psi'\to J/\psi\pi^0)}{{\cal
B}(\psi'\to J/\psi\eta)} =
{27\over16R^2}\left|{\vec{q}_\pi\over\vec{q}_\eta}\right|^3(1+\Delta_{\psi'}),
\label{eq:usratio}
\ee%
where $\vec{q}_{\pi(\eta)}$ denotes the momentum of the pion (eta) in
the rest frame of the $\psi'$ and $\Delta_{\psi'}$ represents SU(3)
breaking effects. Assuming 
$\Delta_{\psi'}<0.4$, an upper limit of $R$ was determined through
Eq.~(\ref{eq:udratio})~\cite{Leutwyler:1996qg}. It can also be
obtained by constructing a chiral effective Lagrangian for
charmonium states and light mesons in a
soft-exchange-approximation~\cite{Casalbuoni:1992fd}. The amplitude
for the $\psi'\to J/\psi\pi^0$ scales as
 \be%
{\cal M}(\psi'\to J/\psi\pi^0) \sim (m_d-m_u)
\left|\vec{q}_\pi\right|. \label{eq:Mpi0qcdme}
\ee%
Using the relation between the masses of quarks and
mesons~\cite{Weinberg:1977hb,Gasser:1982ap}, Eq.~(\ref{eq:usratio})
may be rewritten as~\cite{Ioffe:1980mx} \be%
R_{\pi^0/\eta} = 3 \left(\frac{m_d-m_u}{m_d+m_u}\right)^2
\frac{F_\pi^2}{F_\eta^2} \frac{M_\pi^4}{M_\eta^4}
\left|{\vec{q}_\pi\over\vec{q}_\eta}\right|^3, \label{eq:udratio} \ee%
where $F_{\pi(\eta)}$ and $M_{\pi(\eta)}$ are the decay constant and
mass of the pion (eta), respectively. Using Eq.~(\ref{eq:udratio}) and
the most recent measurement of the decay-width
ratio~\cite{Mendez:2008kb}
\be%
R_{\pi^0/\eta} = (3.88 \pm 0.23 \pm 0.05)\%, \label{eq:cleo}
\ee%
the up-down quark mass ratio is obtained as \footnote{There is a
discrepancy
  between the CLEO result~\cite{Mendez:2008kb} given in
  Eq.~(\ref{eq:cleo}) and the BES result~\cite{Bai:2004cg},
  $R_{\pi^0/\eta}=(4.8\pm0.5)\%$. If we use the branching fractions of
  the $\psi'\to J/\psi\pi^0$ and $\psi'\to J/\psi\eta$ given by the
  Particle Data Group Ref.~\cite{Amsler:2008zz}, which result in
  $R_{\pi^0/\eta}=(4.0\pm0.3)\%$, the result
  $m_u/m_d = 0.39\pm0.02$ is slightly smaller.}
\be%
 \frac{m_u}{m_d} = 0.40\pm0.01. \label{eq:rudpsi}
\ee%
This value is much smaller than the result obtained from the
time-honored formula~\cite{Weinberg:1977hb} \be%
\frac{m_u}{m_d} = \frac{M_{K^+}^2-M_{K^0}^2+2M_{\pi^0}^2-M_{\pi^+}^2}
{M_{K^0}^2-M_{K^+}^2+M_{\pi^+}^2} = 0.56, \label{eq:rudmesonmass} \ee%
and it is also smaller than the large $N_c$ bound, $m_u/m_d\gtrsim
1/2$, derived in Refs.~\cite{Leutwyler:1996sa,Leutwyler:1996qg}.
Note that Eq.~(\ref{eq:rudmesonmass}) is very little affected by
higher order corrections.  It is therefore of fundamental interest
to understand theoretically the discrepancy between the values of
up-down quark mass ratio determined from different sources. This
Letter is devoted to show that the $\psi'$ decays into
$J/\psi\pi^0(\eta)$ are not suitable for extracting the quark mass
ratio, and hence the seeming discrepancy between
Eq.~(\ref{eq:rudpsi}) and Eq.~(\ref{eq:rudmesonmass}) is
meaningless. The reason underlying this statement is that the
earlier analysis neglected effects from intermediate (virtual)
charmed mesons. Those loops were shown to be important in some
charmonium decays in phenomenological models, see, for instance,
Refs.~\cite{Li:2007xr,Liu:2009dr,Zhang:2009kr}. As we will show,
based on a power counting argument in the spirit of heavy quark
effective field theory (HQEFT), which is supported by an explicit
calculation, these contributions overwhelm the one directly related
to the quark masses.

\begin{figure}[t]
\begin{center}
\vglue-0mm
\includegraphics[width=0.48\textwidth]{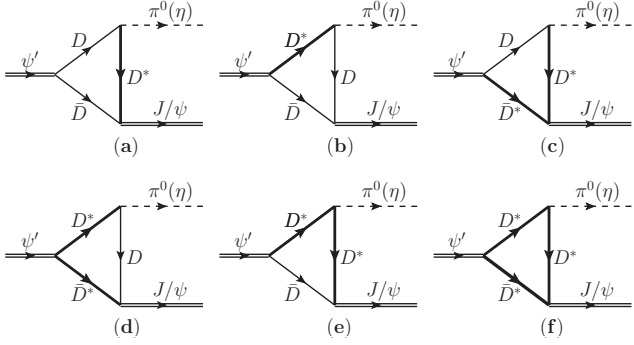}
\vglue-0mm \caption{The decays $\psi'\to J/\psi\pi^0(\eta)$ through
triangle charmed-meson loops. Charmonia, light mesons, pseudoscalar
and vector charmed mesons, are denoted by double, dashed,
thin and solid lines,respectively.
\label{fig:loops}}
\end{center}
\end{figure}

To be specific, we calculate the pertinent diagrams for the decays
$\psi'\to J/\psi\pi^0(\eta)$ involving the lowest lying pseudoscalar
and vector charmed mesons, see Fig.~\ref{fig:loops}. The couplings
of pion and eta to the charmed mesons follow from heavy quark
symmetry and chiral
symmetry~\cite{Burdman:1992gh,Wise:1992hn,Yan:1992gz}. In the
two-component notation of Ref.~\cite{Hu:2005gf}, the charmed mesons
are represented by $H_a=\vec{V}_a\cdot\vec{\sigma}+P_a$ with $V_a$
and $P_a$ denoting the vector and pseudoscalar charmed mesons,
respectively, $\vec{\sigma}$ is the Pauli matrix, and $a$ is the
flavor index. The lowest order axial coupling Lagrangian
is~\cite{Hu:2005gf}
\be%
{\cal L_\phi} = -{g\over 2}{\rm
Tr}\left[H_a^{\dag}H_b\vec{\sigma}\cdot\vec{u}_{ba}\right],
\label{eq:Lphi}
\ee%
where the axial current is
$\vec{u}=-\sqrt{2}\vec{\partial}\phi/F+{\cal O}(\phi^3)$. $F$ is the
pion decay constant in the chiral limit, and the $3\times3$ matrix
$\phi$ collects the octet Goldstone bosons. The leading order
Lagrangian for the coupling of the $J/\psi$ to the charmed and
anti-charmed mesons can be constructed considering parity, charge
parity and spin symmetry. In two-component language, it is
\be%
{\cal L}_\psi = i {g_2\over2}{\rm Tr}\left[J^\dag H_a
\vec{\sigma}\cdot\!\stackrel\leftrightarrow{\partial}\!{\bar
H}_a\right] + {\rm H.c.}, \label{eq:Lpsi}
\ee%
with $A\!\stackrel\leftrightarrow{\partial}\!\!B\equiv
A(\vec{\partial}B)-(\vec{\partial}A)B$. The charmonium field is
given by $J=\vec{\psi}\cdot\vec{\sigma}+\eta_c$ with $\vec{\psi}$
and $\eta_c$ annihilating the $\psi$ and $\eta_c$ states, and ${\bar
H}_a=-\vec{{\bar V}}_a\cdot\vec{\sigma}+{\bar P}_a$ is the field for
anti-charmed mesons~\cite{Fleming:2008yn}. This Lagrangian was first
introduced in Ref.~\cite{Colangelo:2003sa} in four-component
notation with the same coupling $g_2$. Since $\psi'$ is the first
radial excitation of the $J/\psi$, the Lagrangian for the $\psi'$
coupling to the charmed and anti-charmed mesons has the same form as
Eq.~(\ref{eq:Lpsi}) with the coupling constant $g_2$ replaced by the
one for $\psi'$, $g_2'$.

Because the $\psi'$ and $J/\psi$ are SU(3) singlets, it is obvious
that the decay $\psi'\to J/\psi\pi^0$ violates isospin symmetry, and
the decay $\psi'\to J/\psi\eta$ violates SU(3) flavor
symmetry~\footnote{Here we assume the $\eta$ is in the SU(3) octet,
and the effect of the $\eta-\eta'$ mixing is assumed to be
small~\cite{Casalbuoni:1992fd}.}. Accordingly, the decay amplitudes
reflect the flavor symmetry breaking. Here, all the charmed mesons
in a flavor multiplet can contribute, and it is the mass differences
within the multiplet that generates the isospin or the SU(3)
breaking. Similar effects have been studied in $a_0-f_0$
mixing~\cite{Achasov:1979xc,Hanhart:2007bd}, and the isospin
breaking hadronic decay of the
$D_{s0}^*(2317)$~\cite{Faessler:2007gv,Lutz:2007sk,Guo:2008gp}.
Explicitly, the $\psi'\to J/\psi\pi^0$ decay amplitude is
proportional to the difference of the charged and neutral mesons
loops
\be%
{\cal M}(\psi'\to J/\psi\pi^0) \propto \epsilon^{ijk}q_\pi^i
\varepsilon^j_{\psi'}\varepsilon^k_{J/\psi}(I_c-I_n),
\ee%
where $\varepsilon^j_{J/\psi(\psi')}$ denotes the spatial component
of the polarization vector of the $J/\psi(\psi')$,   $I_c$ and $I_n$
are the loop integral expressions which will be given below in
Eq.~(\ref{eq:I}) for charged and neutral charmed mesons. Denoting
the expression for the strange charmed-meson loop by $I_s$, one
obtains the decay amplitude for the  $\psi'\to J/\psi\eta$
\be%
{\cal M}(\psi'\to J/\psi\eta) \propto \epsilon^{ijk}q_\eta^i
\varepsilon^j_{\psi'}\varepsilon^k_{J/\psi} {1\over \sqrt{3}}
(I_c+I_n-2I_s).
\ee%

Before performing the explicit evaluation of the loops it is
important to first understand the power counting of the system. As
was just derived, each vertex in the triangle diagrams is of
$p$-wave character and is thus linear in the momentum. Due to parity
conservation, one momentum has to appear as external parameter (c.f.
Eq. (\ref{eq:Mpi0qcdme})). Thus the loops themselves scale as
$v^3/(v^2)^2v^2=v$, where we replaced momentum factors by the
dimensionless velocities --- the proper expansion parameter of HQEFT
--- and the factors denote the non-relativistic integral measure
and propagators as well as the vertex factors just described, in
order. The typical heavy meson velocity in the loops may be
estimated via  $v\sim
\sqrt{(2M_{\hat{D}}-M_{\hat{\psi}})/M_{\hat{D}}}\simeq0.53$, where
$M_{\hat{D}}$ is the averaged charmed-meson mass, and
$M_{\hat{\psi}}=(M_{J/\psi}+M_{\psi'})/2$. The quantities of
interest here are differences of loops with the remaining terms
proportional to $m_q$
--- this is an energy scale of ${\cal O}(v^2)$. We therefore expect the
heavy meson loops to scale as $m_q/v \, |\vec q|$ which gives some
enhancement compared to Eq.~(\ref{eq:Mpi0qcdme}).

To confirm this power counting estimate and allow for a more
quantitative statement, we now evaluate the diagrams of
Fig.~\ref{fig:loops} explicitly using the non-relativistic
technique. Let us consider diagram (b) in Fig.~\ref{fig:loops} as an
example of these calculations. The decay
amplitude in $d$ dimensions is given by%
\vglue-10mm
\begin{widetext}
\ba%
{\cal M}_{(\rm b)} &\!\!=&\!\! 2i{g\over F}g_{\psi DD}g_{\psi'DD^*}
\epsilon^{ijk}q^i \varepsilon^j_{\psi'}\varepsilon^l_{J/\psi}
\frac{\sqrt{M_DM_{D^*}}}{8M_D^2M_{D^*}} \nonumber\\
&\!\!&\!\! \times \int\!\!{d^dl\over (2\pi)^d}
\frac{l^k(2l^l-q^l)}{\left(l^0-{\vec{l}^{\
2}\over2M_{D^*}}+i\epsilon\right) \left(l^0+b'_{DD^*}+{\vec{l}^{\
2}\over2M_{D}}-i\epsilon\right)
\left(l^0-q^0+\Delta_D-{(\vec{l}-\vec{q})^2\over2M_{D}}+i\epsilon\right)}
\nonumber\\
&\!\!=&\!\! -{g\over 2F}g_{\psi DD}g_{\psi'DD^*} \epsilon^{ijk}q^i
\varepsilon^j_{\psi'}\varepsilon^l_{J/\psi}
\frac{\sqrt{M_DM_{D^*}}}{M_D+M_{D^*}} \int_0^1\!\!{dx}
\int\!\!{d^{d-1}l\over (2\pi)^{d-1}}
\frac{l^k(2l^l-q^l)}{\left[\left(\vec{l}-x\vec{q}/2\right)^2 +
\Delta_{\rm (b)} - i\epsilon\right]^2} \nonumber\\
&\!\!=&\!\! {g\over 8\pi F}g_{\psi DD}g_{\psi'DD^*}
\epsilon^{ijk}q^i \varepsilon^j_{\psi'}\varepsilon^k_{J/\psi}
\frac{\sqrt{M_DM_{D^*}}}{M_D+M_{D^*}} \int_0^1\!\!{dx}
\sqrt{\Delta_{\rm (b)}}, \label{eq:Mb}
\ea%
\end{widetext}
where $q$ is the $\pi^0(\eta)$ momentum, $\Delta_D=M_{D^*}-M_D$ is
the mass difference between the vector and the pseudoscalar charmed
mesons, $\Delta_{(\rm b)} = -ax^2+(c-c')x+c'$, and $a=\vec{q}^{\
2}/4$, $c=b_{DD}M_D+\vec{q}^{\ 2}/2$ and $c'=2\mu_{DD^*} b'_{DD^*}$.
Further, $\mu_{DD^*}=M_DM_{D^*}/\left(M_D+M_{D^*}\right)$ is the
reduced mass of the $D$ and $D^*$, $b'_{DD^*}=M_D+M_{D^*}-M_{\psi'}$
and $b_{DD}=2M_D-E_{J/\psi}$ with $E_{J/\psi}$ being the energy of
the $J/\psi$ in the $\psi'$ rest-frame. The amplitude in
Eq.~(\ref{eq:Mb}) has been multiplied by a factor of
$\sqrt{M_{\psi'}M_{J/\psi}M_D^4M_{D^*}^2}$
 to account for the non-relativistic normalization
of the heavy meson fields in the Lagrangians given in
Eqs.~(\ref{eq:Lphi}) and (\ref{eq:Lpsi}). The dimensionless coupling
constants $g_{\psi DD}$ and $g_{\psi'DD^*}$ are related to the
dimensionful ones $g_2$ and $g_2'$ via $g_{\psi
DD}=g_2\sqrt{M_{J/\psi}M_D^2}$ and $g_{\psi'
DD^*}=g_2'\sqrt{M_{\psi'}M_DM_{D^*}}$. Note in the last step, we
have taken $d=4$. The integral is finite when evaluated with
dimensional regularization for only a power divergence appears.
 The
integral appearing in Eq.~(\ref{eq:Mb}) for $c>a>0,c'>0$, which is
satisfied here, is given by
\begin{widetext}
\ba%
\int_0^1dx\sqrt{\Delta_{(\rm b)}} &\!\!=&\!\! {1\over
4a}\left\{2a\sqrt{c-a} + (c-c')\left(\sqrt{c'}-\sqrt{c-a}\right) +
\frac{(c-c')^2+4ac'}{2\sqrt{a}} \right.\nonumber\\
&\times&\left. \arctan\frac{2\sqrt{a}\left[2a\sqrt{c'}+(c-c')(\sqrt{c-a}-\sqrt{c'})\right]}
{(c-c')^2+2a(c'-c)+4a\sqrt{c'(c-a)}}\right\} 
= {2\over3}{c+c'+\sqrt{cc'}\over \sqrt{c}+\sqrt{c'}}\left[1 + {\cal
O}\left({a\over c}\right)\right]. \label{eq:loopb}
\ea%
\end{widetext}
We checked that neglecting the $a$ term, which is proportional to
$\vec{q}^{\ 2}$, only makes a difference of several percent.
Thus, neglecting the ${\cal O}(\vec{q}^{\
2})$ terms, the decay amplitude of any loop shown in
Fig.~\ref{fig:loops} scales as
\be%
I\equiv {2\over3}\frac{2\mu b + 2\mu' b' + \sqrt{2\mu b 2\mu'
b'}}{\sqrt{2\mu b}+\sqrt{2\mu' b'}}, \label{eq:I}
\ee%
where $\mu (\mu')$ is the reduced mass of the charmed mesons
connected to the $J/\psi(\psi')$ in the loop, and $b(b')$ is the
difference between the charmed meson threshold and
$E_{J/\psi}(M_{\psi'})$.

We may now compare the explicit expressions to the power counting
argument presented above.  Since $\sqrt{2\mu b}$ and $\sqrt{2\mu'
b'}$ are approximately the momenta of the charmed mesons in the
loops, we count them as $M_Dv$ with $M_D$ and $v$ being the mass and
velocity of the charmed meson.  For the purpose of the power
counting analysis, one can neglect the difference between
$\sqrt{2\mu b}$ and $\sqrt{2\mu' b'}$. Then one has $I\sim\sqrt{2\mu
b}$. Denoting the charged and neutral charmed meson mass difference
by $\delta$, we have $\mu_c=\mu_n+\delta/2$ and $b_c=b_n+2\delta$
where the lower index $c$ or $n$ means charged or neutral. Thus,
\ba%
{\cal M}(\psi'\to J/\psi\pi^0) &\!\!\sim&\!\!
|\vec{q}_\pi|(\sqrt{2\mu_c
  b_c} - \sqrt{2\mu_n b_n}) \nonumber\\
&\!\!=&\!\! |\vec{q}_\pi|\delta\frac{2\mu_n+b_n/2}{\sqrt{2\mu_n b_n}}
+
{\cal O}(\delta^2) \nonumber\\
&\!\!\sim&\!\! |\vec{q}_\pi|\frac{\delta}{v}. \label{eq:pc} \ea%
The mass difference $\delta$ may be divided into the strong
(quark-mass difference) and the em contributions as
$\lambda(m_d-m_u)+\beta e^2$, see Ref.~\cite{Guo:2008gp}, therefore
${\cal M}(\psi'\to J/\psi\pi^0)$ scales as
$(m_d-m_u)\left|\vec{q}_\pi\right|/v$ in line with the estimate
given above.

The validity of Eq.~(\ref{eq:udratio}) is based on the assumption
that the light mesons are produced through soft gluons, and hence at
a distance much larger than the size of the charmonium, which is the
basic assumption of the QCD multipole
expansion~\cite{qcdme1,qcdme2,qcdme3}. Then the matrix element of
the soft gluon operator between the vacuum and a light meson can be
worked out using the axial anomaly and chiral symmetry. In the
mechanism considered in this Letter, the light mesons are produced
through their coupling to the virtual intermediate charmed mesons.
This kind of mechanism was not included in the QCD multipole
expansion. These contributions are genuine, i.e. there is no
underlying double counting. This can be seen from the fact that the
corresponding integrals are finite in dimensional regularization and
that the leading terms are non-analytic in the quark masses.
Comparing Eq.~(\ref{eq:pc}) with Eq.~(\ref{eq:Mpi0qcdme}), one sees
that the charmed-meson loop effects in the amplitude are enhanced by
a factor of $1/v\sim2$. Therefore, they are more important.

Assuming the intermediate charmed-meson loop mechanism saturates the
decay widths of the $\psi'\to J/\psi\pi^0(\eta)$, we get
\be%
R_{\pi^0/\eta} = 0.14 \pm 0.09,
\ee%
where the central value is what we get from a direct calculation,
and the uncertainty is from neglecting the contribution of
Eq.~(\ref{eq:Mpi0qcdme}) using $v\sim
\sqrt{(2M_{\hat{D}}-M_{\hat{\psi}})/M_{\hat{D}}}\simeq0.53$ and
contains the one that originates from either using physical masses
or averaged masses for the field normalizations. This value is
within 2$\sigma$ of the experimental ratio. Note that in the ratio
all the coupling constants $g,g_2$ and $g_2'$ disappear.

We cannot give a prediction for the corresponding decays
$\Upsilon'\to \Upsilon\pi^0(\eta)$ by naively extending the
formalism to the bottom sector. This is because the strong and the
em contributions to $M_{B^0}-M_{B^+}$ interfere
destructively~\cite{Guo:2008ns}, and make $M_{B^0}-M_{B^+}$ as small
as $0.33\pm0.06$~MeV~\cite{Amsler:2008zz}. Accordingly, although the
$B$-meson loop contribution to the decay $\Upsilon'\to \Upsilon\eta$
is more important than in the charm sector---$v$ is smaller---its
contribution to the $\Upsilon'\to \Upsilon\pi^0$ is highly
suppressed. On the contrary, in the charmed sector of relevance
here, the strong and em contributions to $M_{D^+}-M_{D^0}$ interfere
constructively~\cite{Guo:2008gp}, and hence enhance the meson loop
effects.

Further support of the proposed scheme is provided by analyzing the
resulting absolute values of the decay widths. Using $g=0.6$, which
is extracted from a tree level calculation of the $D^{*+}$ width,
$F=92.4$~MeV, and $g_{\psi D^{(*)}_{(s)}D^{(*)}_{(s)}}=g_{\psi'
D^{(*)}_{(s)}D^{(*)}_{(s)}}=G$, we obtain the absolute values of the
decay widths as
\ba%
\Gamma(\psi'\to J/\psi\pi^0) &\!\!=&\!\!
(3.6\pm1.9~[5.5\pm2.9])\times10^{-4} G^4 ~{\rm keV},
\nonumber\\
\Gamma(\psi'\to J/\psi\eta) &\!\!=&\!\!
(2.5\pm1.3~[4.9\pm2.6])\times10^{-3} G^4 ~{\rm keV},\non\\
\label{eq:widths}
\ea%
where the numbers outside and inside the square brackets are
obtained using the physical and averaged masses for the field
normalizations, respectively. In order to reproduce the experimental
values $\Gamma(\psi'\to J/\psi\pi^0)=0.40\pm0.03$~keV and
$\Gamma(\psi'\to J/\psi\eta)=10.0\pm0.4$~keV~\cite{Amsler:2008zz},
we need $G=6.2\pm0.9~[5.5\pm0.9]$ and $G=8.4\pm1.3~[7.1\pm1.1]$,
respectively. These numbers are close to independent model estimates
for $g_{\psi DD}$ existing in the literature, see e.g.
Refs.~\cite{Colangelo:2003sa,Haglin:2000ar,Deandrea:2003pv}.

In this context, we want to stress that extracting the quark mass
ratio $1/R$ defined in Eq.~(\ref{eq:R}) using $\Gamma(\eta' \to
\pi^0 \pi^+\pi^-)/\Gamma(\eta' \to \eta \pi^+\pi^-)$ has been
questioned in Ref.~\cite{Borasoy:2006uv}, where meson loops were
also shown to be significant.

In summary, in this Letter, utilizing the technique of
non-relativistic effective field theory, we show that intermediate
charmed mesons play an important role in the $\psi'$ decays into
$J/\psi\pi^0$ and $J/\psi\eta$. They are enhanced by a factor of
$1/v$ compared with the contribution directly from the quark mass
differences. The light quark mass ratio can only be extracted from
these decays after establishing a complete effective field theory up
to next-to-leading order with the Goldstone bosons, charmonia and
charmed mesons as the degrees of freedom in the future. What was
done in this Letter can be regarded as the first step towards that
goal.


We would like to thank B. Kubis for helpful discussion. This work is
partially supported by the Helm\-holtz Association through funds
provided to the virtual institute ``Spin and strong QCD''
(VH-VI-231) and by the DFG (SFB/TR 16, ``Subnuclear Structure of
Matter''). We also acknowledge the support of the European
Community-Research Infrastructure Integrating Activity ``Study of
Strongly Interacting Matter'' (acronym HadronPhysics2, Grant
Agreement n. 227431) under the Seventh Framework Programme of EU.

\end{document}